\documentclass[onecolumn,showpacs,preprintnumbers,amsmath,amssymb]{revtex4}

\usepackage{graphicx}
\usepackage{dcolumn}
\usepackage{bm}
\usepackage{textcomp}
\usepackage{wasysym}
\usepackage{stmaryrd}

\def \ee{\end{equation}}
\def \be{\begin{equation}}

\preprint{}

\begin{document}

\title{Geometrizing the Quantum - A Toy Model}

\author{Benjamin Koch}
 \affiliation{
 Pontificia Universidad Cat\'{o}lica de Chile, \\
Av. Vicu\~{n}a Mackenna 4860, \\
Santiago, Chile \\
}
\date{\today}

\begin{abstract}
It is shown that the equations of relativistic Bohmian mechanics
for multiple bosonic particles have a dual description
in terms of a classical theory of conformally ``curved'' space-time.
This shows that it is possible to formulate
quantum mechanics as a purely classical geometrical theory.
The results are further generalized to interactions 
with an external electromagnetic field.
\end{abstract}

\maketitle

\subsection{Introduction}

In standard quantum mechanics observables and the
corresponding uncertainty are promoted to a fundamental
principle. 
But it was shown by David Bohm that this does not
necessarily have to be the case \cite{Bohm:1951}.
In the de Broglie-Bohm (dBB) interpretation 
it is explained that the uncertainty ``principle'' and the
description by means of operators can be understood
in terms of uncontrollable initial conditions and non-local
interactions between an additional field (the ``pilot wave'')
and the measuring apparatus. 
This theory was further generalized to relativistic quantum mechanics
and quantum field theory with bosonic and fermionic fields
\cite{Holland:1985ud,Nikolic:2002mi,Struyve:2006cj,Nikolic:2006az}.
It is well known that (due to its non-locality) the dBB theory is not
in contradiction to the Bell inequalities \cite{Bell:1988}.
Due to its contextuality, the dBB theory is also not
affected by the Kochen-Specker theorem \cite{Kochen:1976}.

One mayor drawback of the dBB theory is that the
pilot wave and the corresponding ``quantum potential''
have to be imposed by hand without further justification.
In a previous work \cite{Koch:2008hn} 
it was shown that the 
relativistic dBB theory for a single particle is dual
to a scalar theory of curved spacetime.
In this dual theory the ominous ``pilot wave'' can be readily
interpreted as a well known physical quantity, namely a space-time
dependent conformal factor of the metric.
This work on the single particle has many features
in common with other publications on the subject
\cite{Santamato:1984qe,Shojai:2000us,Bonal:2000zc,Carroll:2004hs,Carroll:2007zh,
Abraham:2008yr}.
However, having a duality for the single particle case
is not enough, because the dBB interpretation only is
a consistent quantum theory when it also has the many particle case. The many
particle theory is for example crucial for understanding the
quantum uncertainty and for evading the non existence theorems
\cite{Bell:1988,Kochen:1976}.
Therefore, we will generalize the previous results and 
present a dual for the relativistic many particle dBB theory. 
%
\subsection{Relativistic dBB for many particles}

In this section we shortly list the ingredients for
the interpretation of the many particle quantum Klein-Gordon equation
in terms of Bohmian trajectories. 
For a detailed description of subsequent topics in the dBB 
theory like particle creation,
the theory of quantum measurement,
many particle states, and quantum field theory
the reader is referred to \cite{Nikolic:2006az}.
Let $|0\rangle$ be the vacuum and $|n\rangle$ be an arbitrary n-particle state.
The corresponding n-particle wave function is \cite{Nikolic:2002mi}
$
\psi(x_1;\;\dots\; ; x_n)= \frac{{\mathcal{P}}_s}{\sqrt{n!}}
\langle 0|\hat{\Phi}(x_1) \dots \hat{\Phi}(x_n)|n\rangle$,
where the $\hat{\Phi}(x_j)$ are scalar Klein-Gordon field operators 
and the symbol ${\mathcal{P}}_s$ denotes symmetrization over all 
positions $x_j$.
For free fields, the wave function
satisfies the equation
\be\label{eq_KG0}
\left(\sum_j^n \partial^m_j\partial_{j m}+n \frac{M^2}{\hbar^2} \right)
\psi(x_1;\;\dots\; ; x_n)=0 \quad.
\ee
The mass of a single particle is given by $M$.
The index $j$ indicates which one of the $n$ particles
is affected and the index $m$ is the typical space-time index
in four flat dimensions. 
The wave function allows further for the construction of a conserved
current
$
\sum_j^n \partial^m_j( i \psi^* {\partial}_{j m}
\psi-i \psi {\partial}_{j m}\psi^*)=0$.
This is still standard quantum mechanics, in order to
arrive at the dBB interpretation one rewrites the wave function
$\psi = P e^{iS/\hbar}$ by introducing a real amplitude 
$P(x_1;\;\dots\; ; x_n)$ and a real phase $S(x_1;\;\dots\; ; x_n)$.
Doing this, the complex equation (\ref{eq_KG0}) splits
up into two real equations
\begin{eqnarray} 
\label{eq_KG1}
2 M Q\equiv\sum_j^n(\partial^m_j S)(\partial_{jm} S)
-n M^2,\quad \mbox{with}&Q=\sum_j^n\frac{\hbar^2}{2M} \frac{\partial^m_j \partial_{jm} P}{P}\\
\label{eq_KG2}
0\equiv
\sum_j^n\partial_{j m} \left(P^2 (\partial^m_j S) \right)\quad, 
\end{eqnarray}
where $Q(x_1;\;\dots\; ; x_n)$ is  the quantum potential, and $P(x_1;\;\dots\;
;x_n)$ is the pilot wave.
The first equation can be interpreted as a classical Hamilton-Jacobi
equation with the additional potential $Q$ and
the second equation takes the form of a conserved current.
This is the only way that the $\hbar$ enters
into the dBB theory.
In the dBB interpretation one postulates the existence
of particle trajectories $x_j^m(s)$ whose momentum
$p_j^m$ satisfies the relation
\be\label{eq_KG3}
p^m_j = M\frac{d x_j^m}{ds}\equiv-\partial^m_j S \quad.
\ee
Now one can derive this expression with respect to $ds$
and use the identity
\be\label{eq_der1}
\frac{d}{ds}=\sum_j^n \frac{dx^m_j}{ds}\partial_{jm} \quad.
\ee
This gives the equation of motion for all $n$ relativistic particles 
in the dBB interpretation
\be\label{eq_eom}
\frac{d^2x_j^m}{ds^2}=\sum_i^n\frac{(\partial^l_iS)(\partial_j^m\partial_{il}
S)}{M^2}\quad.
\ee
By using equation (\ref{eq_KG1}) this can be further simplified to 
\be\label{eq_eom1}
M\frac{d^2x^m_j}{ds^2}=\partial^m_j Q \quad.
\ee
The infinitesimal parameter $s$ is not
necessarily time, because every single particle
carries its own reference frame. 
For convenience one might try to choose $s$
as the eigen-time of the particle which is
finally subject to a measurement.
The non-local nature of the dBB theory
becomes obvious in the above equations of motion.
The trajectory of the particle $j$ is determined
from the potential $Q(x_1,\dots, x_n)$, which depends
on the positions of all other particles of the system.

The equations (\ref{eq_KG2}-\ref{eq_eom}) are the building 
blocks of the many particle dBB theory.
The functions $P$, $S$, and $Q$ that appear in those equations
depend on the $4\times n$ coordinates $x_j^m$.
It is therefore possible to introduce a single $4n$ dimensional
coordinate 
$x^L=(x_1^0,x_1^1,x_1^2,x_1^3;\;\dots\;;x_n^0,x_n^1,x_n^2,x_n^3)$, 
which has a capital Latin index and contains the space-time positions of all $n$
particles.
One further observes that in all equations every summation
over a particle index $j$ is accompanied by a summation
over the space-time index $m$ of the corresponding particle. This allows
to replace $\partial_{jm}\rightarrow \partial_L$
and $\partial_{j}^m\rightarrow \partial^L$.
Thus, one can rewrite the equations for the
many particle case (\ref{eq_KG1}-\ref{eq_eom}) as
\begin{eqnarray}
\label{eq_KG1L}
2 M Q\equiv(\partial^L S)(\partial_{L} S)-n M^2\quad \mbox{with}&
Q\equiv\frac{\hbar^2}{2M} \frac{\partial^L \partial_L P}{P},\\
\label{eq_KG2L}
0\equiv
\partial_{L} \left(P^2 (\partial^L S) \right)\quad, \\ 
\label{eq_KG3L}
p^L \equiv M\frac{d x^L}{ds}\equiv-\partial^L S \quad, \\ 
\label{eq_eomL}
\frac{d^2x^L}{ds^2}=\frac{
(\partial^N S)(\partial^L\partial_N
S)}{M^2}\quad\mbox{with}&\frac{d}{ds}\equiv \frac{dx^L}{ds}\partial_{L} \quad.
\end{eqnarray}

\subsection{A $4\times n$ dimensional toy model} \label{sec4n}
\unboldmath

We will now show that the equations of the
many particle dBB theory (\ref{eq_KG1L}-\ref{eq_eomL})
have a dual description in a scalar theory of
curved space-time in $4\times n$ dimensions.
As generalization of the previous single particle approach
\cite{Koch:2008hn} we will
define a setup where the momentum of every particle
is defined in the particles own four dimensional space-time.
Such a $n$-particle theory is therefore
defined in a $4\times n$ dimensional space-time.
Following the flat notation
the coordinates in curved space-time will be denoted as
$\hat{x}^\Lambda=
(\hat{x}_1^0,\hat{x}_1^1,\hat{x}_1^2,\hat{x}_1^3;\;\dots\;;\hat{
x } _n^0 , \hat{x}_n^1 , \hat{x}_n^2 , \hat{x}_n^3)$.

The theory can be formulated by starting from a 
n-particle equation of motion
\begin{eqnarray}\label{eq_equation}
{\mathcal{P}}_s[\hat{R}]=
{\mathcal{P}}_s[\hat{T}]\quad.
\end{eqnarray}
Here, $\hat{R}$ is the Ricci scalar, $\hat{{\mathcal{L}}}_M$ is the 
stress energy tensor of matter, and
$\kappa$ is the coupling constant of this theory.
${\mathcal{P}}_s$ is a symmetrization operator
between different particles $x_i^\lambda$ and $x_j^\lambda$.
This means that
that the fields that are a solution of the problem have to be symmetric
under exchange of two particle coordinates $x_i^\lambda$
and $x_j^\lambda$. It also means
that all the different particles agree on their definition
of what is the $x_j^\mu$ direction. Therefore,
if one wants to perform a coordinate transformation
in a single four dimensional subspace, one has to perform
the same transformation in all other four dimensional subspaces.
In order to describe the local conformal part of this theory
separately one splits the metric $\hat{g}$ up into a conformal
function $\phi(x)$ and a non-conformal part $g$
\be\label{eq_gmn}
\hat{g}_{\Lambda \Gamma}=\phi^{\frac{2}{2n-1}} g_{L G}\quad.
\ee
The index notation for tensors used 
here is explained in table (\ref{tab_IC}) and
it allows to write the equations in $D=4\times n$ dimensions
either with one index (capital letters), 
or with two indices (lower case letters).
A further distinction is made between indices that
are shifted by the metric $\hat{g}$ (Greek)
and indices that are shifted by the metric $g$ (Latin).
\renewcommand\arraystretch{1.5}
\begin{table}[h]
\begin{tabular}{|c|c|c|}
\hline
Index notation & Shifted by $\hat{g}$ & Shifted by $g$ \\
\hline 
Single index & $\hat{X}_\Lambda; \dots | \{\Lambda :1 \dots
4n\} $ & $X_L; \dots |
\{L :1 \dots 4n\}$ \\
\hline
Double index & $\hat{X}_{i \lambda};\dots | \{i:1\dots \,n\,\,\}$ &
$X_{il};\dots |
\{i:1\dots \,n\,\,\}$ \\ \hline
\end{tabular}
\caption{\label{tab_IC} Index convention}
\end{table}
\renewcommand\arraystretch{1.0}
In this section we will only use the single index
notation, but all results can be immediately translated
into the double index notation used in the previous section.
The inverse of the metric (\ref{eq_gmn}) is
$\hat{g}^{\Lambda \Gamma}=\phi^{-\frac{2}{2n-1}} g^{L G}$.
Indices with a lower
Greek and a lower Roman index can be identified
$\hat{\partial}_\Lambda\equiv \partial_L$.
From this follows for example that the adjoint derivatives are not identical,
in both notations
\be\label{eq_derivatives}
\hat{\partial}^\Lambda=\hat{g}^{\Lambda \Sigma}
\hat{\partial}_\Sigma=\phi^{-\frac{2}{2n-1}}
g^{L S}\partial_S=
\phi^{-\frac{2}{2n-1}}\partial^L\quad.
\ee

The geometrical dual to the first dBB equation:
The definition of the metric (\ref{eq_gmn})
allows to reformulate the model
in terms of the separate functions $\phi$ and $g_{L D}$
\cite{HelayelNeto:1999tm}.
Keeping in mind the symmetrization condition
the equation (\ref{eq_equation}) reads
$\frac{2(4n-1)}{1-2n} (\partial^L \partial_L \phi)= \phi(-R+\kappa
{\mathcal{T}}_M)$.
We are primarily interested in studying
a flat Minkowski background space-time $g_{LD}=\eta_{LD}$.
This can be achieved by adding an additional condition
term to (\ref{eq_equation}) that demands a vanishing 
Weyl curvature.
Such a condition also appears in the scalar theories of
curved space-time suggested by Gunnar Nordstr{\"o}m
\cite{Nordstrom:1913a}.
Like in standard general relativity
one further imposes that the metric has a vanishing covariant derivative.
From those conditions it follows
that the metric $g_{LG}$ has only $\pm 1$ on the diagonal, 
while all other entries vanish
$g_{LG}=\eta_{LG}$.
Thus, $R=0$, which
simplifies the above equation to
\begin{eqnarray}\label{eq_equation3}
\frac{2(4n-1)}{1-2n} (\partial^L \partial_L \phi)= \kappa \phi
{\mathcal{T}}_M\;.
\end{eqnarray}
An Extension of the Hamilton Jacobi stress energy tensor
$\hat{{\mathcal{T}}}_M$ can be defined 
by subtracting a mass term
$\hat{M}^2$ for every particle on finds
\begin{eqnarray}\label{eq_TM}
\hat{\mathcal{T}}_M &=& \hat{p}^\Lambda \hat{p}_\Lambda - 
n \hat{M}_G^2\\ \nonumber
&=&\phi^{\frac{-2}{2n-1}}\left(({\partial}^{L}S_H)({\partial}_{L}
S_H)-n M_G^2\right)\quad.
\end{eqnarray}
The Hamilton principle function $S_H$ defines the local momentum
$\hat{p}^\Lambda=\hat{M_G}
\,d\hat{x}^\Lambda/d\hat{s}=-\hat{\partial}^\Lambda
S_H$.
Plugging this into equation (\ref{eq_equation3}) gives
\be\label{eq_Nord1b}
\frac{2(4n-1)}{\kappa(1-2n)}\frac{\partial^L\partial_L \phi}{\phi}=
 ({\partial}^{L}S_H)({\partial}_{L}
S_H)-n M_G^2
\ee
Now one can see that this is exactly the first dBB equation (\ref{eq_KG1L})
if one identifies
\begin{eqnarray}\label{eq_match}
\phi(x)=P(x)\quad, & S_H(x)= S(x)\;\;, \\
\kappa=\frac{2(4n-1)}{1-2n}/\hbar^2\;\;, &
M^2=M_G^2\quad. \nonumber
\end{eqnarray}
Note that the matching conditions demand like in the single
particle case \cite{Koch:2008hn} a negative coupling $\kappa$.

The geometrical dual to the third dBB equation:
According to the Hamilton-Jacobi formalism
the derivatives of the Hamilton principle function ($S_H$)
define the momenta 
\be\label{eq_dualKG3L}
\hat{p}_\Lambda\equiv -(\hat{\partial}_\Lambda S_H)\quad.
\ee
Therefore, with the prescription (\ref{eq_derivatives}) and
the matching condition (\ref{eq_match}) one sees immediately
that the third Bohmian equation (\ref{eq_KG3L})
is fulfilled.

The geometrical dual to the second dBB equation:
In order to find the dual to the second Bohmian
equation one has to exploit that the stress-energy tensor (\ref{eq_TM})
is covariantly conserved
$\hat{\nabla}_\Lambda \hat{T}^{\Lambda \Delta}=0$.
This is true if the following relations
are fulfilled
\begin{eqnarray}\label{eq_cons1}
\hat{\nabla}_\Lambda (\hat{\partial}^\Lambda S_H)=0,&
(\hat{\partial}^\Lambda S_H) 
\hat{\nabla}^\Delta (\hat{\partial}_\Lambda S_H)=0, &
(\hat{\partial}^\Lambda S_H) 
\hat{\nabla}_\Lambda (\hat{\partial}^\Delta S_H)=0.
\end{eqnarray}
Now
one needs to know the 
Levi Civita connection
\begin{eqnarray}\label{eq_Levi}
\Gamma^\Sigma_{\Lambda \Delta}&=&\frac{1}{2}
\phi^{-\frac{2}{2n-1}}\left[(\partial_L 
\phi^{\frac{2}{2n-1}})\delta^S_D
+(\partial_D  \phi^{\frac{2}{2n-1}})\delta^S_L-(\partial^S 
\phi^{\frac{2}{2n-1}})\eta_{LD}
\right]\quad.
\end{eqnarray}
With this, the condition (\ref{eq_cons1}) reads
\be\label{eq_dualKG2L}
\hat{\nabla}_\Lambda (\hat{\partial}^\Lambda S_H)=
\phi^{-\frac{4n}{2n-1}}\partial_L \left[
\phi^2 (\partial^L S_H)
\right]=0\quad.
\ee
With the matching conditions (\ref{eq_match}), 
the above equation is identical to the second
Bohmian equation (\ref{eq_KG2L}).

The geometrical dual to the dBB equation of motion:
The total derivative (\ref{eq_der1}) is generalized to
$\frac{d}{d\hat{s}}=\frac{d\hat{x}^\Lambda}{d\hat{s}}\hat{\partial}_\Lambda=
\phi^{2/(1-2n)} \frac{dx^L}{ds}\partial_L=
\phi^{2/(1-2n)}\frac{d}{ds}$.
Applying this to the momentum 
 $\hat{M}_G (d\hat{x}^\Lambda/d\hat{s})=(\hat{\partial}^\Lambda S_H)$
gives the equation of motion
$\hat{M}\frac{d^2 \hat{x}^\Lambda}{d\hat{s}^2}=
\frac{1}{\hat{M}}
(\hat{\partial}^\Delta S_H)\hat{\partial}_\Delta(\hat{\partial}^\Lambda
S_H)$,
or equivalently the equation of motion in the Minkowski coordinates $x^L$
\be\label{eq_eom22}
\frac{d^2x^L}{ds^2}=\frac{
(\partial^N S_H)(\partial^L\partial_N
S_H)}{M^2} \quad.
\ee
Using (\ref{eq_match}) one sees that the equation of motion (\ref{eq_eom22}) 
is dual to the equation of motion of relativistic Bohmian
mechanics (\ref{eq_eomL}). 
This is however almost a triviality because the
equations (\ref{eq_eomL}) and (\ref{eq_eom22}) are
all derived from the same mathematical prescription (\ref{eq_der1}).
But in curved space-time there is an other equation of motion
in addition to the equation of motion (\ref{eq_eom22}), the 
geodesic equation which can be be shown to lead to the same result.
%
\subsubsection{Interaction with an external field}
Now, the
results of the previous sections will be generalized
to interactions with an external electromagnetic field.
%

An external field in the dBB theory:
Coupling $n$ bosonic particles (\ref{eq_KG0}) 
with charge $e$ to an external 
electromagnetic field $A_m$ is achieved by
replacing the partial derivative 
with a gauge covariant derivative
$\partial_m \rightarrow \partial_m + i e A_m/\hbar$
in the Klein-Gordon Lagrangian.
The resulting equation of motion is
\be\label{eq_KG0c}
\sum_j^n\left[ \left(
\partial^m_j\partial_{j m}+\frac{M}{\hbar^2}-
 \frac{e^2 A_j^2}{\hbar^2}\right)\psi+ 
\frac{i\partial^m_j (\psi^2 A_{jm})}{\hbar}
\right]
=0\;,
\ee
where $A_j^m$ is the electromagnetic potential $A^m$ evaluated
at the position of particle $j$.
By again rewriting the $n$-particle
wave function $\psi=P \exp (iS)/\hbar$ the two
equations (\ref{eq_KG1}, \ref{eq_KG2}) generalize to
\begin{eqnarray} 
\label{eq_KG1c}
2MQ&\equiv&\sum_j^n(\partial^m_j S+eA_j^m)(\partial_{jm} S+eA_{jm})
-n M^2 ,\;\;\;\\
\label{eq_KG2c}
0&\equiv&
\sum_j^n\partial_{j m} \left(P^2 (\partial^m_j S+eA^m_j) \right)\quad.
\end{eqnarray}
In the presence of an external force, the
dBB definition for the particles momentum (\ref{eq_KG3})
now contains the canonical momentum $\pi_j^m$ instead
of the normal momentum $p_j^m$
\be\label{eq_KG3c}
\pi^m_j = M\frac{d x_j^m}{d\tau}\equiv-(\partial^m_j S+e A_j^m) \quad.
\ee
Thus, using (\ref{eq_der1}, \ref{eq_KG1c}, \ref{eq_KG3c},
and the relation $\partial_{kn}A_j^m=\delta_{kj} \partial_n A_j^m$)
one finds the equation of motion for all $n$ particles
in an external field $A^m$
\be\label{eq_eom1c}
M\frac{d^2x^m_j}{ds^2}=\partial^m_j Q + e
 \pi_{j n} F^{mn}\quad.
\ee
Here, the field strength tensor is
$F^{mn}_{jk}=\partial^m_j A^n_k-\partial^n_k A^m_j=\delta_{jk}F^{mn}$ 
with $F^{mn}=\partial^m A^n-\partial^n A^m$.
On the RHS of equation (\ref{eq_eom1c}) appear two terms.
The first term is the quantum potential also present in equation (\ref{eq_eom1})
and the second term is the Lorentz force which is
familiar from classical electrodynamics.
Now one can undo the formal rewriting of coordinates
and the interacting equations (\ref{eq_KG1c}-\ref{eq_eom1c})
read
\begin{eqnarray}
\label{eq_KG1Lc}
2 M Q&\equiv&(\partial^L S+eA^L)(\partial_{L}S+eA_L)-n M^2\quad \\
\label{eq_KG2Lc}
0&\equiv&
\partial_{L} \left(P^2 (\partial^L S+eA^L) \right)\quad, \\ 
\label{eq_KG3Lc}
p^L& \equiv& M\frac{d x^L}{ds}\equiv-(\partial^L S+eA^L) \quad, \\ 
\label{eq_eomLc}
M\frac{d^2x^L}{ds^2}&=&\partial^L Q +e \pi_K F^{LK} \quad.
\end{eqnarray}

An external field in the $4\times n$ dimensional theory of curved
space-time:
In the interacting case, the classical
$4 \times n$ dimensional theory of curved space-time
is analogous to the discussion in section \ref{sec4n}.
The only difference appears 
in the definition of the canonical momentum
$\hat{\pi}^\Lambda=\hat{M}_G \frac{d \hat{x}^\Lambda}{d\hat{s}}
= - (\hat{\partial}^\Lambda S_H + e \hat{A}^\Lambda) $,
instead of the free momentum $\hat{p}^\Lambda$.
With this replacement the equations 
(\ref{eq_Nord1b},\ref{eq_dualKG2L},and \ref{eq_dualKG3L})
transform to
\begin{eqnarray}
\label{eq_dualKG1Lc}
\frac{2(4n-1)}{\kappa(1-2n)}\frac{\partial^L\partial_L\phi}{\phi}
&\equiv&(\partial^L S_H+eA^L)(\partial_{L}S_H+eA_L)-nM^2_G\quad,\\
\label{eq_dualKG2Lc}
0&\equiv&
\phi^{\frac{4n}{1-2n}}\partial_{L} \left(P^2 (\partial^L S_H+eA^L)
\right)\,  \quad.
\end{eqnarray}
One immediately sees that with the identifications
(\ref{eq_match}) the above equations 
are dual to the equations (\ref{eq_KG1Lc}-\ref{eq_KG3Lc}) 
In order to check the equation of motion (\ref{eq_eomLc})
we use the total derivative and find
$\frac{d^2\hat{x}^\Lambda}{d\hat{s}^2}=
\phi^{\frac{4}{1-2n}}\frac{d^2 x^L}{ds^2}
+\hat{\pi}^L \cdot (\dots).
$
Using this and 
$\label{eq_cons3c}
\hat{\pi}^\Lambda \hat{\nabla}_\Lambda \hat{\pi}^\Delta =0$
one can verify that the geodesic equation
is consistent with (\ref{eq_eomLc}).

\subsubsection{Discussions}
In this section some interpretational 
issues are addressed. This aims to develop a physical
understanding of the presented duality.

Locality:
Quantum mechanics in the dBB interpretation
is a theory which allows to talk about particle
position at the cost of non-local interactions.
The non-locality becomes obvious 
by looking at the equation of motion
for a free particle in the dBB theory (\ref{eq_eom1}).
The movement of one particle is governed by
the quantum potential $Q$, which simultaneously
depends on the positions of all the other
particles of the system. 
The next question is:
``How can a non-local theory have a dual local theory?''
In the presented theory, every single particle is living in
its own four dimensional space-time. Therefore,
the positions of $n$ different particles correspond
to one single point in the $4\times n$ dimensional
space-time. 
This higher dimensional construction helps around
the non-locality argument, but it has the 
price that the theory has to be formulated
in the $4\times n$ configuration space.

Time:
One notes the appearance of $n$ time coordinates in this formulation.
However, this possible conceptual problem can be evaded 
since the actual quantum observables are restricted by
equal time commutation relations. They form a subset of the
solutions allowed here. One might therefore only consider
solutions where all $x_i^0$ are equal.
In other approaches that differ from this one 
it is put forward that the nature of quantum
mechanics could be a consequence of an additional time 
dimension \cite{Koch:2008hn,Foster:2010vt}.

Gravity:
``How is this geometrical theory related to THE geometrical theory
- general relativity?''
It is tempting to speculate about 
a generalization of the given toy model 
to tensor equations.
However, writing the analogous higher dimensional tensor
equations leads to a large set of strongly coupled
differential equations. It will have
to be explored whether in some limit, classical general relativity
and the dBB theory are part of the
space of solutions.

The matching conditions:
The matching conditions (\ref{eq_match})
are not unique. They were chosen
in order to have a direct connection
between three functions.
This has the consequence
that if one gives $\hbar$ a fixed numerical
value, then the coupling of the geometrical
theory $\kappa$ runs from $-6/\hbar^2$ to $-4/\hbar^2$, depending
on the number of particles ($n=1, \dots ,\infty$).
Reversely demanding a fixed geometrical coupling
would result in a running Planck constant $\hbar$.
One might see the scale dependence
of the coupling $\kappa$ as a feature of the
toy model which it shares with effective quantum
field theories and even some approaches to gravity \cite{Reuter:2001ag}.

\subsection{Summary}

It was  shown
that the equations of the free relativistic dBB theory
for many particles (\ref{eq_KG1L}-\ref{eq_eomL})
have a dual description in a  geometrical theory in $4\times n$ dimensions.
For the translation between the two theories a single
set of matching conditions was defined (\ref{eq_match}).
The result was then generalized to interactions with an
external electromagnetic field.
The question whether such dualities can
also be found for, self-interacting theories,
fermions, or quantum field theory will be subject of future
studies.\\
The author thanks 
M. A. Diaz,
J.M. Isidro,
 and H. Nikolic, 
for helpful comments.



\end{document}